\documentclass[aps,onecolumn,superscriptaddress,showpacs]{revtex4}
\usepackage{graphicx}

\usepackage{subfigure}

\begin{document}

\author{Fabrizio Capuani}

\affiliation{FOM Institute for Atomic and Molecular Physics (AMOLF), Kruislaan 407, 1098 SJ Amsterdam, The Netherlands}

\email{capuani@amolf.nl, frenkel@amolf.nl}

\author{Ignacio Pagonabarraga}

\affiliation{Departament de F\'{\i}sica Fonamental, C. Mart\'{\i} i Franqu\'es 1, 08028 Barcelona, Spain}

\email{ipagonabrraga@ub.edu}

\author{Daan Frenkel}

\affiliation{FOM Institute for Atomic and Molecular Physics (AMOLF), Kruislaan 407, 1098 SJ Amsterdam, The Netherlands}

\email{frenkel@amolf.nl}

\bibliographystyle{unsrt}

\title{Lattice-Boltzmmann simulations of the sedimentation of charged disks}

\pacs{82.70.Dd, 
      82.70.-y, 
      66.10.-x, 
      47.65.+a, 
      82.45.-h 
      }

\begin{abstract}
We report a series of Lattice-Boltzmann simulations of the
sedimentation velocity of charged disks.  In these simulations, we
explicitly account for the hydrodynamic and electrostatic forces
on disks and on their electrical double layer.

By comparing our results with those for spheres with equal surface
and charge, we can clarify the effect of the particle shape on the
sedimentation process. We find that disks and spheres exhibit a
different dependence of the sedimentation velocity on the Debye
screening length. An analysis of the behavior of highly charged
disks (beyond the scope of the linearized Poisson-Boltzmann
equation) shows that, in that regime,  the charge dependence of
the sedimentation velocity of disks and spheres is similar. This
 suggests that,  at high charge, the effective hydrodynamic shape of the disks becomes more spherical.
\end{abstract}

\maketitle

\section{Introduction}
Suspensions of charged disks are of great practical importance.
Examples range from clay suspensions to blood.   In the present
paper, we present calculations of the electrokinetic behavior of
charged disks. As disks are not spherically symmetric, they also
provide an ideal model system to study the effect of  shape on the
coupling between electrostatic and hydrodynamic response of a
macroscopic particle.

There exists an extensive experimental literature on the transport
properties of suspensions of charged disks. Yet, in spite of the
importance of these systems, there is surprisingly little
theoretical knowledge about the effect of the charge of the disks
on their transport properties. One reason may be that the
non-spherical geometry greatly complicates the use of the
analytical approach that is used to describe charged, spherical
particles. Whilst there are papers that consider the
hydrodynamical properties of uncharged disks~\cite{OdlT03} or the
electrostatic properties of charged disks~\cite{KHS+00,ORG04},
we are not aware of any theoretical publications that treat the
interplay between electrostatics and hydrodynamics for charged
disks.

Several theoretical studies suggest that, in general, there may be
non-trivial coupling effects due to shape asymmetries
\cite{AjdariElectrolyte,AjdariOnsager}. The need for a numerical
(rather than an analytical) approach is related to the fact that
the analytical approaches are usually only tractable in certain
limits. For example, to arrive at tractable analytical expressions
it is often necessary to assume that the Poisson-Boltzmann
equation can be linearized or that the Debye screening length is
small compared to the linear dimensions of the charged colloid.
Yet these conditions are rarely satisfied for  charged biocolloids
and biopolymers. These (non-spherical) particles usually have high
charges under physiological conditions making the use of the
linearized Poisson-Boltzmann equation questionable.

The total number of ionizable groups of colloids can be determined
experimentally by titration. However, the experimental
determination of the effective charge of colloids usually exploits
 non-equilibrium techniques, such as electrophoresis or
sedimentation. The interpretation of many of these experiments is based
on theoretical expressions that have been derived for weakly
charged, spherical colloids. Yet, it is not at all obvious that
expressions valid for spheres can be extrapolated to disks. Hence,
it is important to understand the effect of shape and charge on
the transport behavior of non-spherical charged particles.

In the present paper we report a numerical study of the
electrokinetics of charged disks. In order to perform these
simulations, we extended a Lattice-Boltzmann scheme that we had
developed to study the electrokinetic behavior of spherical
colloids. With this simulation method, we can study the dynamics
of charged particles at the Poisson-Boltzmann level. Hence our
simulations ignore charge fluctuations and static charge
correlation effects. In the absence of polyvalent ions, the
Poisson-Boltzmann description is expected to work quite well.

Our simulations extend from the regime where the
electrohydrodynamic coupling is small (low charge, thin double
layer) to the case where the charge is large and the double layer
is extended. By performing simulations of many different
conditions, we disentangle the effects of charge, double layer
width and shape.

The paper is organized as follows. In
Sec.~\ref{sec:Electrokinetic-model} we present the basic
electrokinetic equations and the simulation method that we use to
solve them. In order to study the effect of charge on the
sedimentation of disks, we must first know the sedimentation
behavior of uncharged disks. This is done in Sec.~\ref{sec:3},
where we use our model to compute the friction coefficients of a
neutral sedimenting disk for motions parallel to and perpendicular
to its axis of symmetry (we refer to these motions as
``longitudinal'' and ``transverse'', respectively). In
Sec.~\ref{sec:Charge effect} we discuss the effect of the charge
of the disk on the sedimentation velocity. In Sec.~\ref{Sec:
volume fraction Z, high phi} we consider the effect of the
concentration of the disks on the sedimentation velocity.
Sec.~\ref{sec:diffuse layer} focuses on the variation of the
sedimentation velocity with Debye-screening length. Sec.~\ref{sec:
Shape} contains a comparison of the sedimentation behavior of
disks and spheres. Concluding remarks are contained in
Sec.~\ref{sec:Discussion}.

\section{Electrokinetic model\label{sec:Electrokinetic-model}}

We analyze a simple geometry in which one disk-like colloidal
particle with radius $a$ and height $h$ sediments due to the
action of a uniform external field. The disk has an overall charge
$Q=Ze$, where $Z$ is the valency and $e$ is the elementary charge unit. The aspect ratio $p$ is defined as $p=2a/h\,.$ The disk is
suspended in a symmetric electrolyte and, for the sake of
simplicity, we assume that coions and counterions have the same
mobility. The fluid mixture is characterized locally by the solvent density, $\rho_{s}$, and by the microion electrolyte densities, $\rho_{\pm}.$ The latter also determine the local
charge density of the fluid,
$q({\textbf{r}})=ze[\rho_{+}({\textbf{r}})-\rho_{-}({\textbf{r}})]$.
We restrict ourselves to monovalent electrolytes, i.e. $z=1$.

On a macroscopic length scale, the dynamics of the system is
governed by the standard electrokinetic
equations~\cite{HunterColloids} that specify the interplay between
the electrical potential, local charge density, electrical currents and
fluid flow:

\begin{eqnarray}
\frac{\partial}{\partial t}\rho_{k} & = & -\nabla\cdot\mathbf{j}_{k}\;\;\; k=+,-\label{smoluchowski}\\
\frac{d}{d t}\left(\rho\textbf{v}\right) & = & \eta\nabla^{2}\left(\rho\textbf{v}\right)-\nabla P+\frac{k_{B}\textrm{T}}{e}q\nabla\Phi.\label{electrokinvel}\end{eqnarray}
\begin{equation}
\mathbf{j}_{k}=-\rho_{k}\textbf{v}+D_{k}\left[\nabla\rho_{k}+z_{k}\rho_{k}\nabla\Phi\right]\label{eq:
Flux Smolu}\end{equation}%
where $\eta$ is the shear viscosity, $P$ is the pressure,
$\mathbf{v}$ is the fluid velocity,
$k_{B}\textrm{T}\equiv\beta^{-1}$ measures the temperature, and
$D_{k}$ stands for the diffusivity of each electrolyte species (which reduce to a single constant for symmetric electrolytes). 
$\hat{\Phi}$ is the electrostatic potential, while
$\,\,\Phi\equiv\hat{\Phi}(k_{B}T/e)$ is an appropriate  dimensionless potential which
 satisfies the Poisson equation
\begin{equation} \nabla^{2}\Phi=-4\pi
l_{B}\left[\sum_{k=\pm}z_{k}\rho_{k}+\rho_{w}\right],\label{poisson}\end{equation}%
where $l_{B}=\beta e^{2}/(4\pi\epsilon)$ is the Bjerrum length,
 $\epsilon=\epsilon_{0}\epsilon_{r}$ denotes the dielectric
constant of the medium, whilst $\rho_{w}$ refers to the charge
density due to embedded solid objects, either colloids of solid
walls.

Equation~(\ref{smoluchowski})  simply expresses a conservation law
and Eqn.~(\ref{eq: Flux Smolu}) is the constitutive equation.
Together with the incompressibility condition
$\nabla\cdot\mathbf{v}=0$, Eqn.~(\ref{electrokinvel}) corresponds
to the Navier-Stokes equations for an incompressible, isothermal
electrolyte. In the presence of external forces (such as the
gravitational field), the corresponding force must be added to the
right-hand side of Eqn.~(\ref{electrokinvel}).

\subsection{Simulation method}
To simulate the sedimentation of charged disks, we used the
lattice-Boltzmann scheme  reported in
Ref.~\cite{myjcp2004}. We showed therein that lattice-Boltzmann
can be used to compute transport coefficients of charged
spherical colloids.   Below, we briefly summarize the main
features of the method and refer the reader to
that reference for further details.

The Lattice-Boltzmann (LB) method is the lattice counterpart of
the Boltzmann equation. It prescribes a dynamical evolution rule
for the distribution function $n_{i}(\mathbf{r},t)$, which
represents the density of particles at the lattice node
$\mathbf{r},$ at the discrete time $t$ and with the discrete
velocity $\mathbf{c}_{i}$. The density-weighted moments of the
local  velocity distribution correspond to the hydrodynamic
fields. In particular,
$\sum_{i}{\textbf{c}}_{i}n_{i}({\textbf{r}},t)=\rho{\bf v}$ is the fluid's	
momentum; it satisfies the Navier-Stokes equation on length and
time scales that are large compared to the lattice spacing and the
LB time step, respectively.

The electrolyte species are simulated by following the diffusion
and convection of the local densities of coions and counterions
described by Eqns.~(\ref{smoluchowski}) and (\ref{eq: Flux Smolu}). This equation is based on
the flux of each species along the links that connect neighboring
nodes, and ensures strict local charge conservation. This local
charge, combined with the corresponding electrostatic potential---computed by a numerical solution of the Poisson equation
(\ref{poisson})---provides the local force that accelerates the
fluid. With this technique, colloidal particles are simply
introduced as surfaces where the collision rules of the
populations of the neighboring nodes are modified to ensure
non-slip boundary conditions~\cite{LaddJFMI}. The link-based
definition of the flux of the electrolyte species leads to a
straightforward implementation of the no-flux boundary condition
for each of the ionic species at solid surfaces. This suppresses
possible charge leakage through the solid walls.

For reasons of computational convenience, we choose the value of
the kinematic  viscosity $\nu=1/6$ (in lattice
units)~\cite{LaddJFMI} and the $\rho_s=1$, as the density unity. 
The external (gravitional) field that induces
sedimentation was  chosen to be $10^{-6},$ a value that is well
inside the linear-response regime. This gravitational field
generates fluid velocities of the order of $10^{-8}$  (again, all
expressed in lattice units). The diffusivity of the electrolyte is set to
$D=0.19$, a value for which spurious diffusion due to lattice
advection  (see Ref.~\cite{myjcp2004}) is negligible. The values
of the diffusivity and the flow velocity correspond to P\'{e}clet
numbers smaller than $10^{-1}.$ In the simulations described in
the subsequent sections we vary the salt concentration  between
$7\times10^{-4}$ and $5\times10^{-3}$ as a way to control the
 electrical double layer thickness.

\section{Sedimentation of neutral disks\label{sec:3}}
Before assessing the role of electrostatics on the sedimentation
of non-spherical particles, we performed LB simulations to compute
the sedimentation velocity of uncharged hard disks. Such reference
calculations are needed because, in contrast to the case of hard
spheres, analytic expressions for the sedimentation velocity of an
isolated hard disk only exist in the limit of infinitely thin
disks~\cite{Williams,Brenner}. We are not aware of analytical
results for disks with finite aspect ratios.

We have simulated the sedimentation of disks with two different
nominal aspect ratios $p=10$ and $p=5$, corresponding to disks of
lateral dimension $h=2,$ and radii $a=10$ and $5$, respectively.
However, these aspect ratios are only approximate: in the LB
approach, the hydrodynamic boundary of a solid particle is 
usually located close to the midpoint of links joining fluid and solid
nodes. In practice, the hydrodynamic shape of an object may differ
slightly from the nominal one. For this reason, we need to
calibrate the shapes of the disks. An unambiguous way to determine
these effective sizes (the hydrodynamic radius and height) would
be to measure the friction coefficients of the particles and
compare the results with the appropriate analytical expression
of an infinitely thin disk.
Great care must be taken when comparing the LB numerical results
for the friction coefficient with results obtained for an isolated
disk. Since we use periodic boundary conditions,
the simulations measure the friction coefficient of a regular
array of particles at volume fraction $\varphi=\pi a^{2}h/L^{3},$
where $L$ is the diameter of  the simulation box. Hence, in order to
extrapolate to  infinite dilution, we must perform  a series of
simulations with increasing box size. The same procedure will also
be used to compute the sedimentation velocity of a charged disk in
the dilute limit.

\subsection{Friction coefficients\label{Sec: neutral volume exp}}
In the following calculations, the reference frame is centered on the disk and the disk is fixed on the lattice. 
Hence, the gravitational force acting on the disk becomes  a body force, with opposite direction, that acts to the fluid.
We then determined the total fluid velocity at steady state, $v_{F}$. As
this velocity is equal and opposite to the sedimentation velocity
of the particle, $U_{d}=-v_{F}$, we can relate the sedimentation
velocity to the friction coefficient through $U_{d}=F_{g}/\xi$,
where $U_{d}$ is the sedimentation velocity of the disk, $\xi$ is
its friction coefficient, and $F_{g}$ is the applied gravitational
force. We computed both the friction coefficient for  motion along
the symmetry axis of the disk and the one for perpendicular
(transverse) motion.

Hashimoto~\cite{Hashimoto} has shown that the friction coefficient
of an array of hard spheres depends, at low volume fractions $\varphi$,
as $(\xi/\xi_{0})^{-1}=1-1.76\varphi^{1/3}+\varphi+O(\varphi^{2}),$
which shows that the initial decrease in the sedimentation velocity
of an array of hard spheres is controlled by the colloid volume fraction. 
\begin{figure}
\begin{center}\includegraphics[clip,width=0.80 \columnwidth]{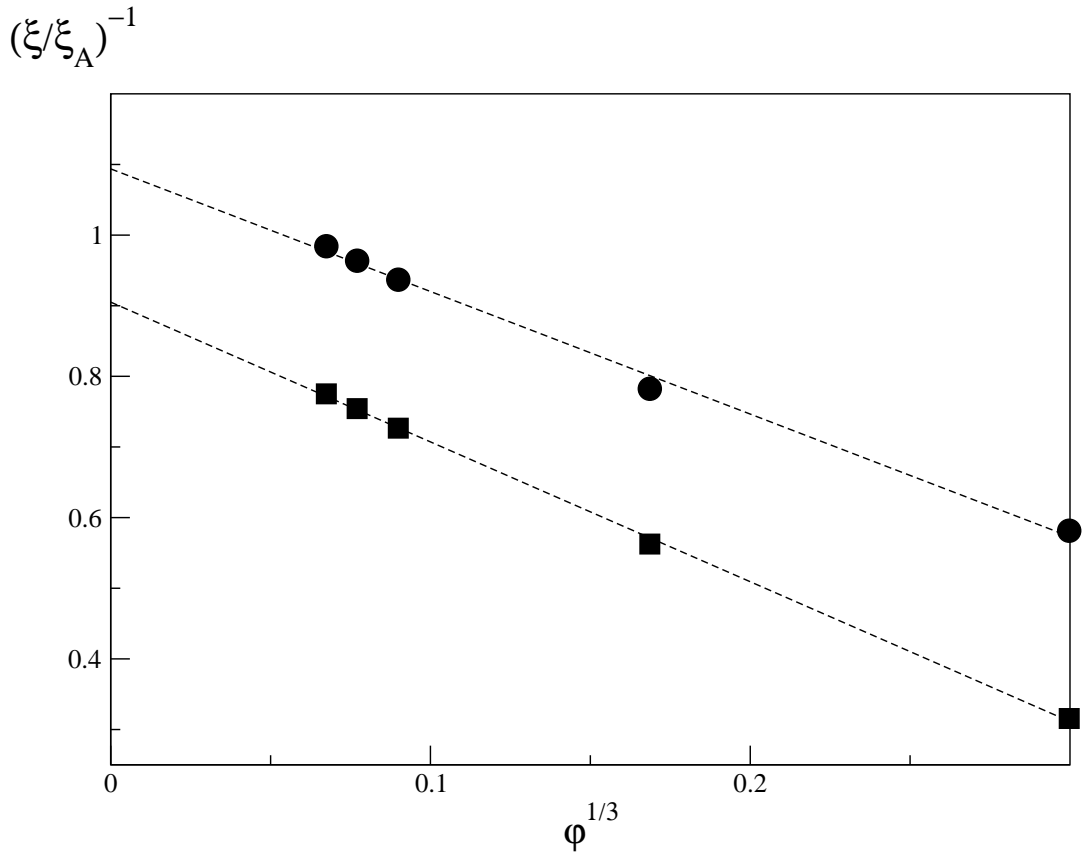}
\end{center}
\caption{Transverse (circles) and longitudinal (squares) friction
coefficients for a sedimenting neutral disk of aspect ratio $p=5$
as function of the  volume fraction of the array. The friction
coefficients are normalized by the friction coefficient of a
sphere with the same radius as the cylinder, i.e.
$\xi_{A}=6\pi\eta R.$ The dashed lines are linear
fits.\label{fig1:neutral friction coefficients}}
\end{figure}

In Figure~\ref{fig1:neutral friction coefficients} we show the
transverse and longitudinal friction coefficients for a neutral
disk with $p=5$ at various volume fractions, normalized by the
Stokes friction coefficient of a sphere with the same area, i.e.
by $\xi_{A}\equiv6\pi\eta R$, $R=\sqrt{a(a+h)/2}.$ The figure
shows that, just as in the case of neutral spheres, the friction
varies linearly with $\varphi^{1/3}$. Knowledge of this
concentration dependence allows us to extrapolate the numerical
results  to estimate the friction
coefficients ($\xi_{\perp}$ and $\xi_{\parallel}$) at infinite
dilution.  One limiting case is known:   
the friction coefficients of an infinitely thin disks ($p\rightarrow\infty$) is identical to that of an oblate spheroid with the same aspect ratios between its main axis~\cite{Brenner}.

We summarize the results for the normalized friction coefficients
for disks of two aspect ratios in Table~\ref{DI Table}. 
For the disk with $p=5$, we conclude that, to a good approximation, the
hydrodynamic radius and height correspond to the nominal ones. The
numerical values obtained for the larger disk ($p=10$) are less
satisfactory than for the shorter one. 
While one could conclude that, for a larger aspect ratio the hydrodynamic radius and height 
are different with respect to the nominal ones, and hence 
recompute the effective values for "$a$" and $h$ so as to adjust $\xi_A$, 
this does not seem satisfactory, since for a larger object the
disagreement between the nominal and hydrodynamic size is expected
to decrease. Moreover, because we have always used cubic simulation boxes,
we can attribute such a difference to the stronger coupling between
image disks for longitudinal sedimentation.
On the other hand, the ratio between the perpendicular and
parallel frictions $\xi_{\perp}/\xi_{\parallel}$ do agree with the
value estimated on the basis of the nominal size, suggesting that
the deviations come mostly from uncertainties related to the
extrapolation from finite volume fraction values. 
Hence, we conclude that, also for
this shape, the disagreement between the two sizes is negligible,
and we ascribe the deviations to interactions with the periodic
images.

\begin{table}
\begin{center}\begin{tabular}{|c|c|c|c|c|}
\hline
$\xi_{A}=6\pi\eta R$&
\multicolumn{2}{c|}{ p=5}&
\multicolumn{2}{c|}{p=10}\tabularnewline
\hline
&
 Simulation&
 Oblate spheroid&
 Simulation&
 Oblate spheroid\tabularnewline
\hline
$\xi_{\perp}/\xi_{A}$&
 0.91&
 0.9&
 0.93&
 0.85\tabularnewline
\hline $\xi_{\parallel}/\xi_{A}$&
 1.1&
 1.06&
 1.19&
 1.11\tabularnewline
\hline $\xi_{\perp}/\xi_{\parallel}$&
 0.83&
 0.84&
 0.78&
 0.77  \tabularnewline
\hline
\end{tabular}\end{center}

\caption{Friction coefficients of an isolated disk in transverse
and in longitudinal motion normalized by the Stokes friction
coefficient $\xi_{A}=6\pi\eta R$ of a sphere with equal surface
area, i.e. with radius $R=\sqrt{a(a+h)/2}$ ($a$ is the radius of
the cylinder). We compare the computed value of the friction
coefficients with approximate theoretical values for an oblate
spheroid with the two axis equal to the disk radius and to the
disk height. We studied two disks with aspect ratios $p=5$ and
$p=10$, respectively. \label{DI Table}}
\end{table}

\section{Sedimentation velocities of charged disks: charge dependence\label{sec:Charge effect}}
Before discussing the computed sedimentation velocities of charged
disks, we briefly recall the theoretical results concerning the
sedimentation velocity of weakly charged spheres, since this
theory serves as a reference point for the discussion of the
results for disks. Booth~\cite{Booth} (and Ohshima \emph{et
al.}~\cite{OhshimaSed}) predicted that the sedimentation velocity $U_{s}(Z)$, where Z is the valence, of
an isolated sphere  is a quadratic function of the
sphere valence, which can be expressed as
\begin{equation}
\frac{U_{s}(Z)}{U_{s}(0)}=1-c_{2}(\kappa R)Z^{2},
\label{eq:booth}
\end{equation}%
where $U_{s}(0)$ corresponds to the sedimentation velocity of a hard sphere in
the dilute limit. In the regime where the Debye-H\"{u}ckel theory
is valid,  the pre-factor $c_{2}$ for a symmetric 1-1 electrolyte
is given by
\begin{equation} 
c_{2}(\kappa R)=\frac{k_{B}Tl_{B}}{72\pi
R^{2}\eta D}f(\kappa R),
\label{booth simple}
\end{equation}%
where $R$ is the radius of the sphere and, with $\rho^0_k$ denoting the bulk value of the charge density of species $k$, $\kappa=\sqrt{4\pi
l_{B}\sum_{k}z_{k}^{2}\rho^0_{k}}$ stands for the inverse Debye length that
characterizes the size of the electrical double layer. The
function $f(\kappa R)$ is defined as 
\begin{eqnarray}
f(\kappa R) & = & \frac{1}{1+(\kappa R)^{2}}\left[\textrm{e}^{2\kappa R}\left(3E_{4}(\kappa R)-5E_{6}(\kappa R)\right)^{2}+8\textrm{e}^{\kappa R}\left(E_{3}(\kappa R)-E_{5}(\kappa R)\right)\right.\nonumber \\
 &  & \left.-\textrm{e}^{2\kappa R}\left(4E_{3}(2\kappa R)+3E_{4}(2\kappa R)-7E_{8}(2\kappa R)\right)\right],
 \label{GE: nonumber1}
 \end{eqnarray}%
expressed as a linear combination of the integral function, $ \textrm{E}_{n}(x)\equiv x^{n-1}\int_{x}^{\infty}dt\; t^{-n}\exp(-t)$.

It is reasonable to assume that, in the case of disks, the
dependence of the sedimentation velocity on colloidal charge in
the Debye-H\"{u}ckel limit has the same functional form as
Eqn.~(\ref{eq:booth}), where all the shape dependence and
hydrodynamic coupling enters through the factor $c_{2}(\kappa R).$
To test this, and to analyze the role of charge on the sedimentation velocity of disks, 
we performed  a series of simulations at  constant Debye screening length and  volume fraction ($\varphi=7.2\times10^{-4}$
for  $p=5$, and $\varphi=2.9\times10^{-3}$ for $p=10$).
Since in many cases of practical interest colloidal particles are highly charged~(see e.g~\cite{TKC01} for the case of disk-like
clay particles) we performed numerical simulations covering a wide range of disk charges. 

\begin{figure}
\begin{center}\includegraphics[%
  clip,
  width=0.80\columnwidth]{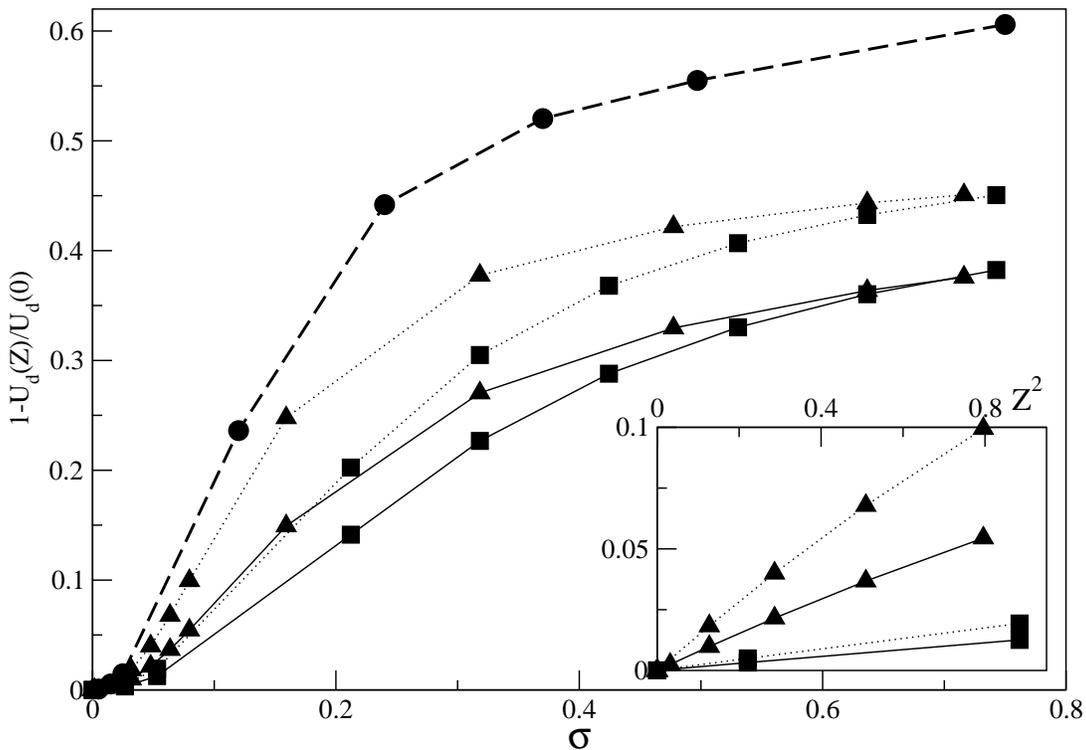}\end{center}

\caption{Surface-charge-density dependence of
$1-U_{d}(Z)/U_{d}(0)$ for a disk with aspect ratio $p=5$ (squares)
and $p=10$ (triangles) in transverse (dotted lines) and
longitudinal (continuum lines) motion at $ka=1.$ For comparison,
we also show the behavior for a sphere with radius $R=4.2$
(circles). In the inset, we show  $1-U_{d}(Z)/U_{d}(0)$
for weakly charged disks as function of $Z^{2}$.  The curves for the disk with
$p=10$ are plotted against $Z^{2}/10$ to show them on the same
scale. The inset shows that for small charges, the sedimentation
velocity varies quadratically with charge. The lines in the inset
are linear fits to the simulation data. In contrast, the curves in
the main figure are simply meant as  a guide to the
eye.\label{fig2: Z dep high charge}}
\end{figure}

The results obtained are displayed in Fig.~\ref{fig2: Z dep high charge}, where the velocity, normalized by the sedimentation velocity of a hard disk at the same volume fraction, is depicted as a function of the surface charge $\sigma=eZ/\left[2\pi a(a+h)\right]$. 
One can identify the quadratic dependence at low charge, consistent with Booth theory for spheres. 
Such a dependence can be clearly appreciated in the figure's inset. 
After this quadratic growth, a crossover region is identified, 
for surfaces charge densities between $0.1$ and $0.4,$ before entering the asymptotic regime at even larger surface charge densities, 
where the sedimentation velocity increases much more slowly.  
This behavior is consistent with numerical results on the sedimentation velocity of charged colloidal spheres 
which also show a deviation from Booth's predictions for surface charge densities around $0.1.$

Comparing the deviation of the sedimentation velocity for
transverse and longitudinal motion, as displayed in
Fig.~\ref{fig2: Z dep high charge}, one can see that the
sedimentation velocity decreases faster with charge for transverse
motion, regardless of the aspect ratio. This is indicative of a
stronger electrokinetic coupling for transverse motion.
Interestingly, the sedimentation velocities (expressed by
$1-U_{d}(Z)/U_{d}(0)$) become almost independent of the aspect
ratio of the disks for large surface-charge densities. In the same
figure, we also plot the decrease in sedimentation velocity for a
sphere of radius $R=4.2$, which shows the same dependence on
surface charge as that of the disks.

Figure~\ref{fig2: Z dep high charge} allows us to draw some
qualitative conclusions concerning the nature of the errors that
are made when estimating the charge of disk-like particles by
assuming the validity of Booth's theory~ \cite{TKC01}. 
As mentioned above, Booth's theory is valid only for \emph{weakly} charged
sedimenting \emph{spheres}. As we now have numerical results for
the sedimentation velocity of disks, we can identify two sources
of errors. At low charge, where the quadratic dependence of the
sedimentation velocity on colloidal charge holds, there will still
be some discrepancy in the coefficient $c_{2}$. We  address this issue
in more detail in Section~\ref{sec: Shape}. At high charge, where
even the quadratic surface-charge dependence does not apply, the
use the Booth theory leads to serious errors in the estimation of
the zeta potential. To give an idea of the magnitude of this
error, one should compare the curves shown in Fig.~\ref{fig2: Z
dep high charge} with a parabolic extrapolation of the curves up
to $\sigma\simeq0.1.$ Our calculations also suggest that if  the
sedimentation velocity of a charged colloid is plotted as a function
of the surface-charge density (and not as a function of the total
charge as one might be tempted to do when comparing with Booth's
theory) the curves corresponding to the sedimentation velocity of
a variety of charged disks show the same functional dependence.

\section{Sedimentation velocities of charged disks: volume fraction dependence\label{Sec: volume fraction Z, high phi}}
As discussed above, the inverse friction coefficient of a dilute,
ordered array of hard spheres scales as $\varphi^{1/3}.$ In
Ref.~\cite{myjcp2004}, we have verified that this functional
dependence also holds for charged spheres, provided that the
system is dilute enough to guarantee that there is no significant
overlap of the double layers. For charged disks, we expect the
$\varphi^{1/3}$ dependence to hold under the same circumstances.
In order to test whether there is a detectable effect of the
overlap of electric double layers of different disks, we have
computed the normalized friction coefficients for disks of aspect
ratio $p=5$ as a function of the volume fraction, for volume
fractions up to $10\%,$ for different widths of the diffuse layer.

In Fig.~\ref{fig:Z10, vol frac dep high PHI} we show the results
for a weakly charged disk, both for transverse and longitudinal
motion. Note that, in the dilute limit, the friction coefficients
will depend on $\kappa a$ due to the electrohydrodynamic
interaction. For transverse motion the convergence to the dilute
limit is slower, indicating a stronger coupling between disks; we
attribute this to the fact that the distance of closest approach
coincides with the external field direction. Although for $\kappa
a=1/2$ and high volume fractions the diffuse layers overlap, the
effect of the diffuse layer is much weaker  than the volume-fraction
dependence or than the effect of the particle shape. The shape
effect is reflected in the substantial difference between the
friction coefficients for transverse and longitudinal
sedimentation. Although barely visible in  Figure~\ref{fig:Z10,
vol frac dep high PHI}, there is a  small but significant
dependence of the friction coefficients on $\kappa$. This we
discuss in more detail in the next section.

The dependence on $\kappa$ is better visible in
Fig.~\ref{fig:Z100, vol frac dep edge and broad}, where we display
the friction coefficients of highly charged disks
Although here the dependence on $\kappa$ is clearly visible, it becomes less
important at higher volume fractions. This suggests that with
increasing  volume fraction, the effect of the overlap of diffuse
layers becomes less important than the direct effect of
hydrodynamic interaction between disks. At small volume fractions,
we always recover the $\varphi^{1/3}$-dependence of the inverse
friction coefficients. However, deviations form this scaling are
already noticeable at small $\varphi$. This fact indicates that
the dilute regime is confined to smaller $\varphi$ values for highly charged disks. In the
limit $\kappa a\rightarrow \infty$ (vanishing Debye length), the
sedimentation friction coefficient of a charged and a neutral
disk should coincide. The fact that, for finite $\kappa a$, the
ratio of the friction coefficients extrapolates at $\varphi=0$ to
a value different from one indicates the importance of
electrokinetic coupling for the sedimentation of charged disks.

We stress that the values of $\kappa$ are computed on the basis of
the electrolyte densities in the bulk. For concentrated
suspensions of highly charged disks, the density of counterions
added to the system to ensure charge neutrality
($\rho_{-}=-Z/V_{\textrm{f}}$, where $Z$ is the valency of the
sphere and $V_{\textrm{f}}$ the volume occupied by the
electrolyte) may exceed the concentration of added salt. In that
case,  $\kappa^{-1}$, the screening length in the salt
``reservoir'' is not simply related to the apparent screening
length in the dense suspension. For our simulations this
phenomenon becomes important only for the highest volume fractions
(typically $\varphi>~0.08$). Hence, our extrapolation at low
$\varphi$   is not affected by this complication.

\begin{figure}
\begin{center}\includegraphics[%
  clip,
  width=0.80\columnwidth]{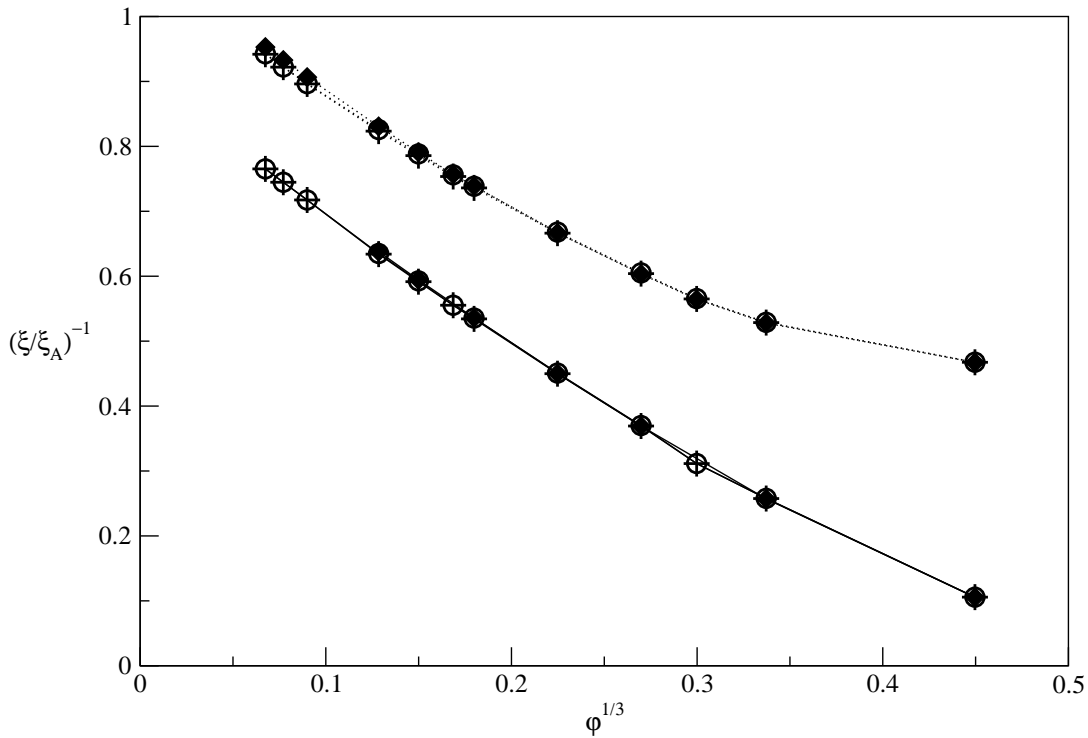}\end{center}

\caption{Volume-fraction-dependent normalized friction
coefficients for a disk with radius $a=5,$ aspect ratio $p=5,$ and
valency $Z=10$ for $\kappa a$ equal to $0.5$ (spheres), $0.8$
(pluses), and $2.1$(filled diamonds). The upper curves are for the
transverse friction coefficient, while the lower curves are for a
longitudinal friction coefficient. The curves are normalized by
the friction coefficient of an isolated sphere with equal surface
area $\xi_{A}=6\pi\eta\sqrt{a(a+h)/2}$ (see Table \ref{DI Table}
for the correspondent neutral values). Curves are drawn as a guide
to the eye.}

\label{fig:Z10, vol frac dep high PHI}
\end{figure}

\begin{figure}[t]
\begin{center}\includegraphics[%
  clip,
  width=0.80\columnwidth]{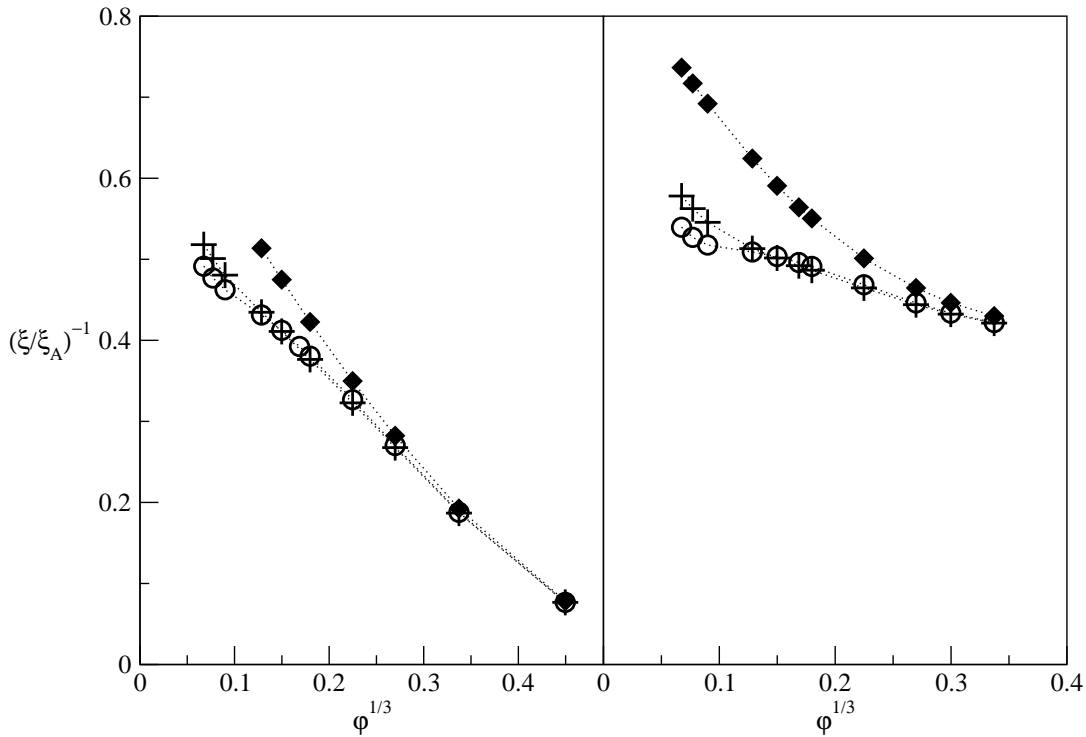}\end{center}

\caption{Volume-fraction-dependent normalized friction
coefficients for a disk with radius $a=5$ and aspect ratio $p=5$
and valency $Z=100$ or $\kappa a$ equal to $0.5$ (spheres), $0.8$
(pluses), and $2.1$(filled diamonds). a) Transverse friction
coefficient; b) Longitudinal friction coefficient. In all
cases,the curves are normalized by the friction coefficient of a
sphere with equal surface area $\xi_{A}=6\pi\eta\sqrt{a(a+h)/2}$
(see Table \ref{DI Table} for the correspondent neutral values).
Curves are drawn as a guide to the eye.}

\label{fig:Z100, vol frac dep edge and broad}
\end{figure}

\section{Sedimentation velocity of charged disks: Debye length dependence\label{sec:diffuse layer}}
Having analyzed the role of charge and volume fraction on the
sedimentation velocity, we now consider in more detail the effect
of the double layer width on the sedimentation of the
disks. We follow the same procedure as in Section~\ref{Sec:
neutral volume exp} and study the sedimentation velocity
$U_{d}(Z)$ at different volume fractions, normalized by the
corresponding velocity of isolated charged-neutral disks
$U_{d}(0)$.
In this way, the ratio $U_{d}(Z)/U_{d}(0)$ measures
the reduction is the sedimentation velocity of one charged disk
due to its electrokinetic interaction with the electrolyte. The
reduction in sedimentation velocity in the dilute limit is
interesting theoretically, because we can compare with analytic
results for weakly charged spheres, although in experiments the
reduced sedimentation velocity $U_{d}(Z,\varphi)/U_{d}(0,\varphi)$
at finite $\varphi$ is the relevant quantity. For simplicity, in
the remaining part of this section, we will be writing $U_{d}(Z)$
instead of $U_{d}(Z,\varphi)$ but, unless explicitly stated, the
volume fraction dependence is always assumed.

\begin{figure}
\begin{center}\subfigure[Transverse motion]{\includegraphics[%
  clip,
  width=0.60\columnwidth]{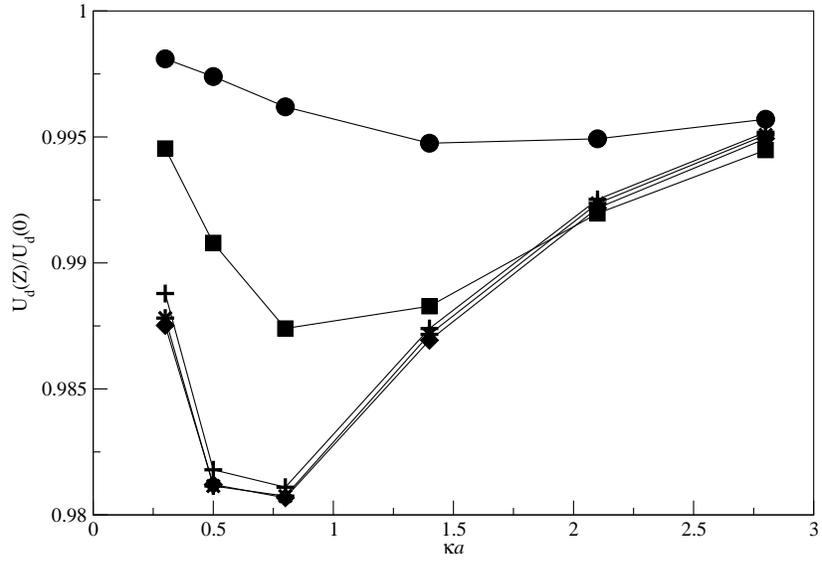}}\end{center}

\begin{center}\subfigure[Longitudinal motion]{\includegraphics[%
  clip,
  width=0.60\columnwidth]{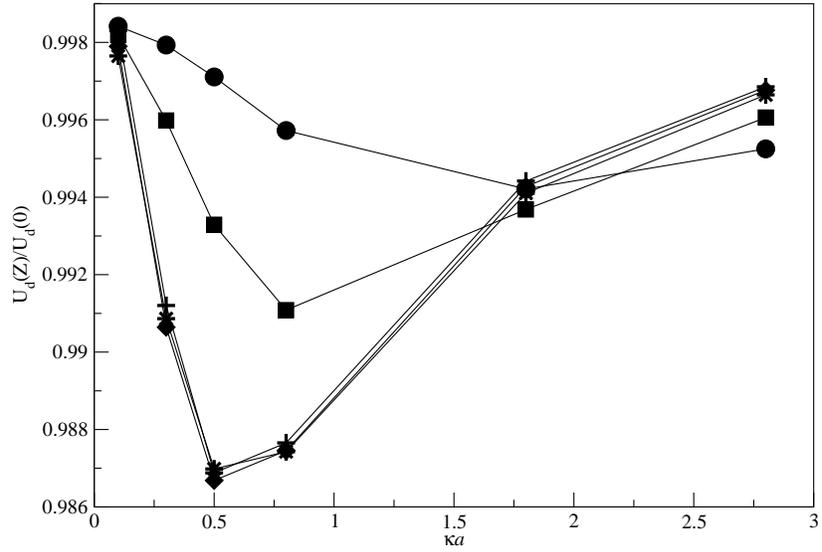}}\end{center}

\caption{Normalized sedimentation velocity of a charged disk with
aspect ratio $p=5,$ radius $a=5,$ and valency $Z=10$ as function of the EDL thickness
expressed in dimensionless units $\kappa a$.
The different simulation points correspond to $\varphi=$
$3.8\times10^{-2}$ (circles), $4.8\times10^{-3}$ (squares),
$7.3\times10^{-4}$ (pluses), $4.6\times10^{-4}$ (stars), and
$3.1\times10^{-4}$ (diamonds). (a) Transverse motion; (b)
Longitudinal motion. Curves are drawn as a guide to the eye.
\label{fig: U/U0 -vs- ka -vs v.f.}}
\end{figure}

In Figure~\ref{fig: U/U0 -vs- ka -vs v.f.}(a), we show the
normalized sedimentation velocity as a function of the double
layer width for a disk with aspect ratio $p=5$ in transverse
motion, for different volume fractions. For infinitely thin and
infinitely broad diffuse layers, the sedimentation velocity should
coincide with that of a charged-neutral disk, and hence the curve
should approach one for both small and large $\kappa a$, as is
indeed observed. The decrease at intermediate values of $\kappa a$
is the result of the interplay between hydrodynamic dissipation
and electrolyte diffusion. The largest effect is observed when the
size of the double layer is of the order of the largest dimension
of the disk, i.e. $\kappa a\sim1$. The effect increases with
decreasing volume fraction, consistent with the discussion in the
previous section, and above volume fractions around $1\%\,,$ the
changes in normalized sedimentation become negligible. The minimum
velocity also depends on volume fraction, an effect which is
consistent with previous findings for spheres~\cite{KD00}. A
similar behavior is observed in Fig.~\ref{fig: U/U0 -vs- ka -vs
v.f.}(b), where longitudinal sedimentation for a weakly charged
disk is depicted. It is interesting to note that the decrease in
sedimentation velocity is slightly smaller. We can ascribe this
effect to the fact that the distorted double layer is not
isotropic and has a smaller contribution to the friction when the
wider side of the disk is exposed to a region where the velocity
gradients are smaller.%

In Figures~\ref{fig: U/U0 transverse Z=3D10, p=3D10}(a) and
\ref{fig: U/U0 transverse Z=3D10, p=3D10}(b), we show the
sedimentation velocity for a weakly charged disk with a somewhat
larger aspect ratio: $p=10$. The trends are the same as for the
disk with $p=5$, although the minimum velocity seems to depend on
aspect ratio, and is achieved now for slightly narrower double
layers. The reduction in absolute terms is now smaller, but this
is simply due to the lower surface charge density as compared with
the smaller ($p=5$) disk. It is worth mentioning that for disks
with $p=5$ we could effectively reach the dilute limit for
$U_{d}(Z)/U_{d}(0)$. In contrast,  for  disks with $p=10$ we had
to perform the dilute-limit extrapolation.

\begin{figure}
\begin{center}\includegraphics[%
  clip,
  width=0.80\columnwidth]{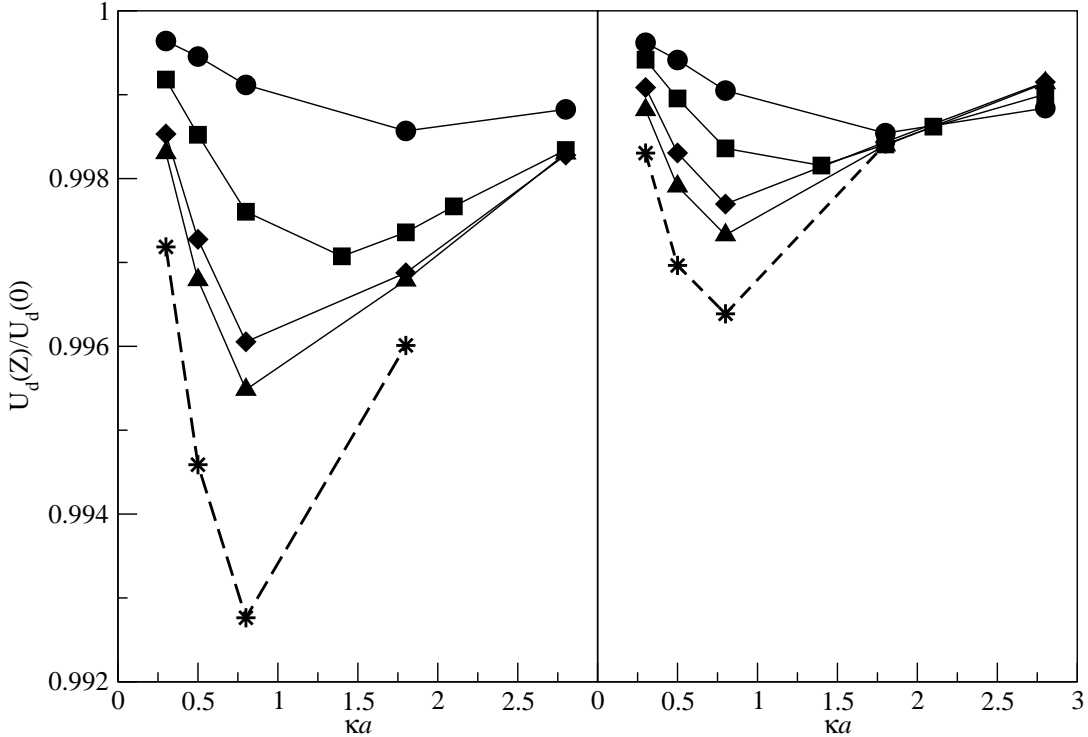}\end{center}

\caption{Normalized sedimentation velocity of a charged disk with
aspect ratio $p=10,$ radius $a=10,$ and valency $Z=10$
as function of the EDL thickness
expressed in dimensionless units $\kappa a$ at volume fractions
$\varphi=$ $1.9\times10^{-2}$ (circles),
$\varphi=6.5\times10^{-3}$ (squares), $\varphi=2.9\times10^{-3}$
(diamonds), and $\varphi=1.8\times10^{-3}$ (triangles), and the
correspondent dilute limit extrapolation (stars). Left: Transverse
motion; right: Longitudinal motion. Curves are drawn as a guide to
the eye. \label{fig: U/U0 transverse Z=3D10, p=3D10}}
\end{figure}
\begin{figure}
\begin{center}\subfigure[Transverse motion]{\includegraphics[%
  clip,
  width=0.60\columnwidth]{Figures/Ud_Ud0_a=5h=2-vs-phi-vs-ka_Z100_PAR}}\end{center}

\begin{center}\subfigure[Braodside motion]{\includegraphics[%
  clip,
  width=0.60\columnwidth]{Figures/Ud_Ud0_a=5h=2-vs-phi-vs-ka_Z100_PER}}\end{center}

\caption{Normalized sedimentation velocity of a charged disk with
aspect ratio $p=5,$ radius $a=5,$ and valency $Z=100$
as function of the EDL thickness
expressed in dimensionless units $\kappa a$
at various volume fractions: $\varphi=$
$3.8\times10^{-2}$ (circles), $4.8\times10^{-3}$ (squares),
$7.3\times10^{-4}$ (pluses), $4.6\times10^{-4}$ (stars), and
$3.1\times10^{-4}$ (diamonds). (a)~Transverse sedimentation; (b)~
longitudinal sedimentation. Curves are drawn as a guide to the
eye. \label{fig: p=3D5 Z100 transverse various volfrac}}
\end{figure}
\begin{figure}
\begin{center}\includegraphics[%
  clip,
  width=0.80\columnwidth]{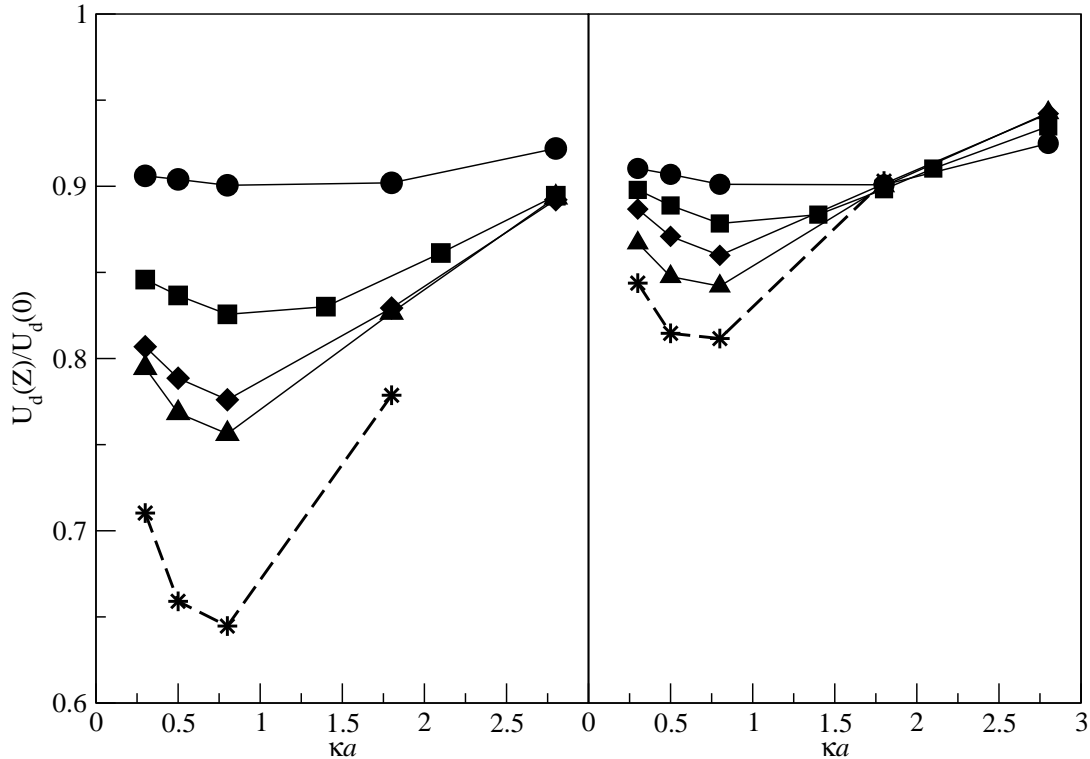}\end{center}

\caption{Volume-fraction-dependent reduction in the sedimentation
velocity of a charged disk with aspect ratio $p=10,$ radius
$a=10,$ and valency $Z=100$. The plot on the left refers to a disk
sedimenting along its edge, while the right part along its
symmetry axis. The different curves correspond to volume fractions
$\varphi=$ $1.9\times10^{-2}$ (circles), $6.5\times10^{-3}$
(squares), $2.9\times10^{-3}$ (diamonds), $1.8\times10^{-3}$
(triangles), and the corresponding dilute limit extrapolation
(dashed line with star symbols). Curves are drawn as a guide to
the eye. \label{fig8: Volume-fraction-dependent-reduction-in}}
\end{figure}

In Figure~\ref{fig: p=3D5 Z100 transverse various volfrac} we show
the sedimentation velocity for a highly charged disk with $p=5$,
normalized by the sedimentation velocity of uncharged disks at the
same volume fraction. Although, again, the relevance of the
electrokinetic coupling in the sedimentation velocity diminishes
upon increasing volume fraction, the coupling between electric
friction and velocity dissipation becomes much more dominant now.
The sedimentation velocities decrease by almost $50\%$, and the
range of the values of $\kappa a$ where appreciable deviations from
the uncharged disk behavior is observed is wider than in
the case of weakly charged disks. Hence, the electrokinetic
coupling for disks is much larger than that observed for
spheres, and is in fact consistent with mobility reductions
observed in laponites\cite{TKC01}. For higher aspect ratios the same trends are observed, as shown in
 Fig.~\ref{fig8: Volume-fraction-dependent-reduction-in}  for the particular aspect ratio $p=10$.

\begin{figure}
\begin{center}\subfigure[Z=10]{\includegraphics[width=0.50\columnwidth,angle=270,clip=true]{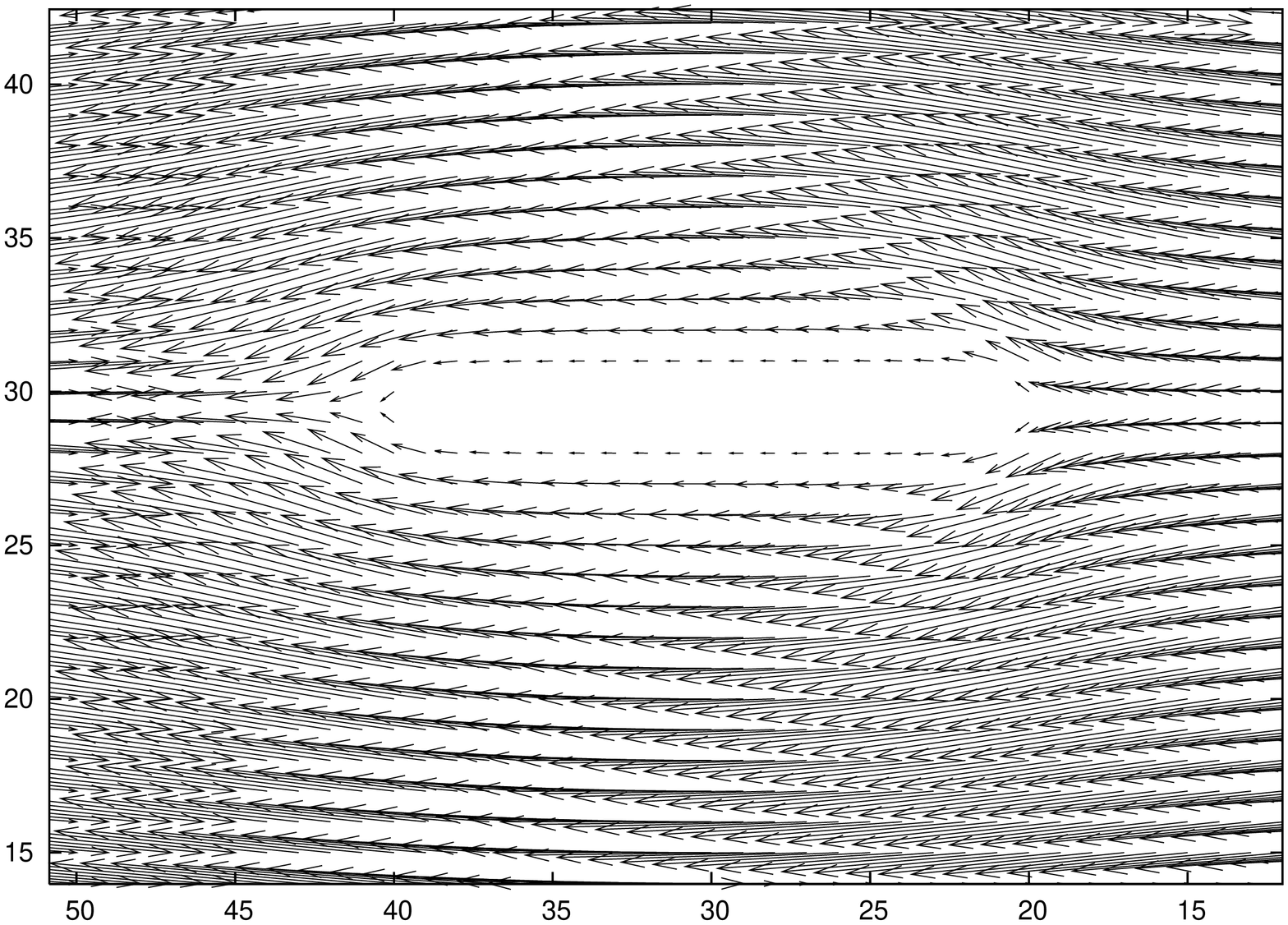}} \subfigure[Z=300]{\includegraphics[width=0.50\columnwidth,angle=270,clip=true]{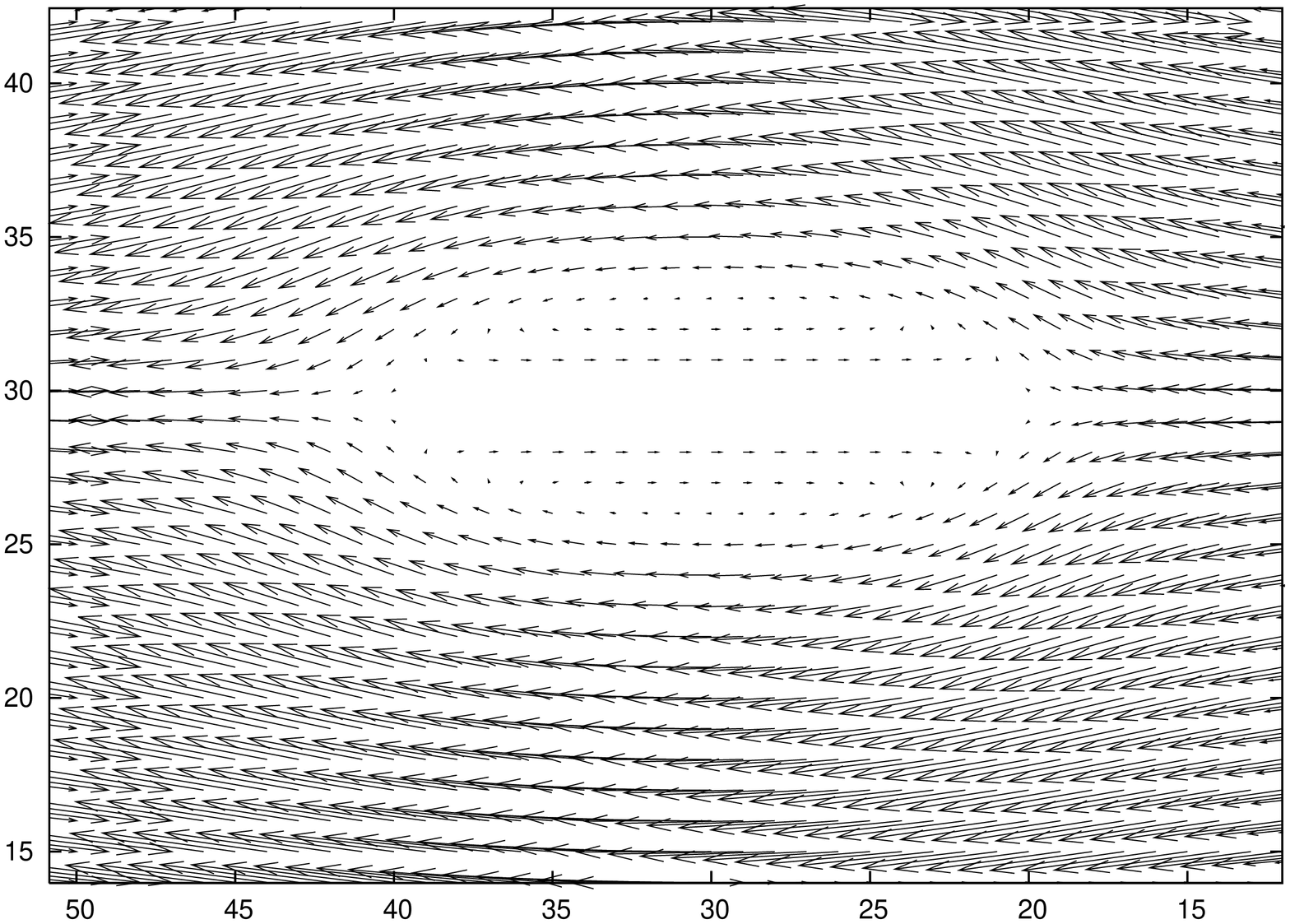}}\end{center}

\caption{Projection of the fluid velocity field near a sedimenting charged disk on
the plane parallel to the axis of revolution of the disk.
The disk sediment along its \emph{edge}.
\label{cap:Flow-edge-side}}
\end{figure}

\begin{figure}
\begin{center}\subfigure[Z=10]{\includegraphics[width=0.50\columnwidth,angle=270,clip=true]{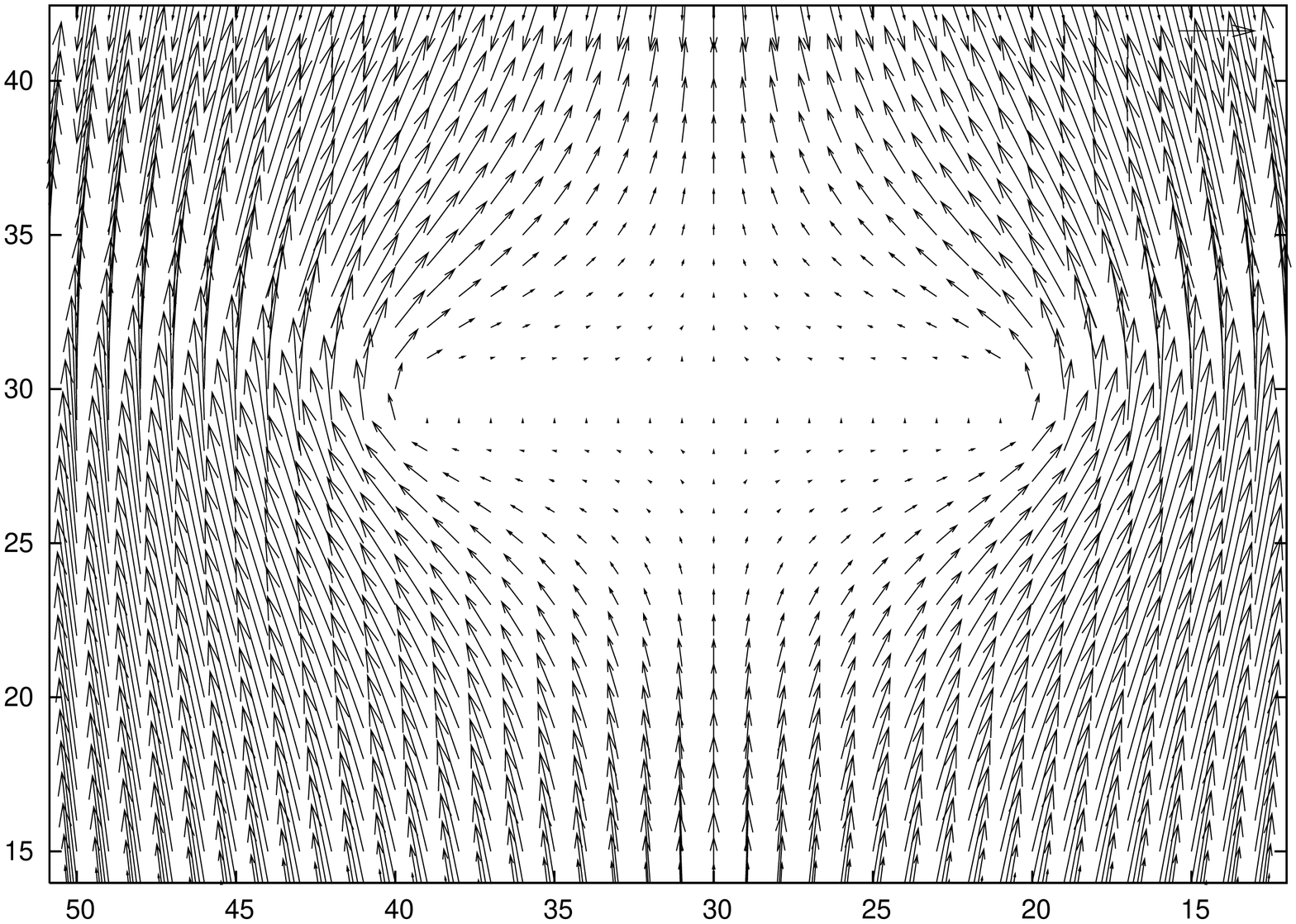}} \subfigure[Z=300]{\includegraphics[width=0.50\columnwidth,angle=270,clip=true]{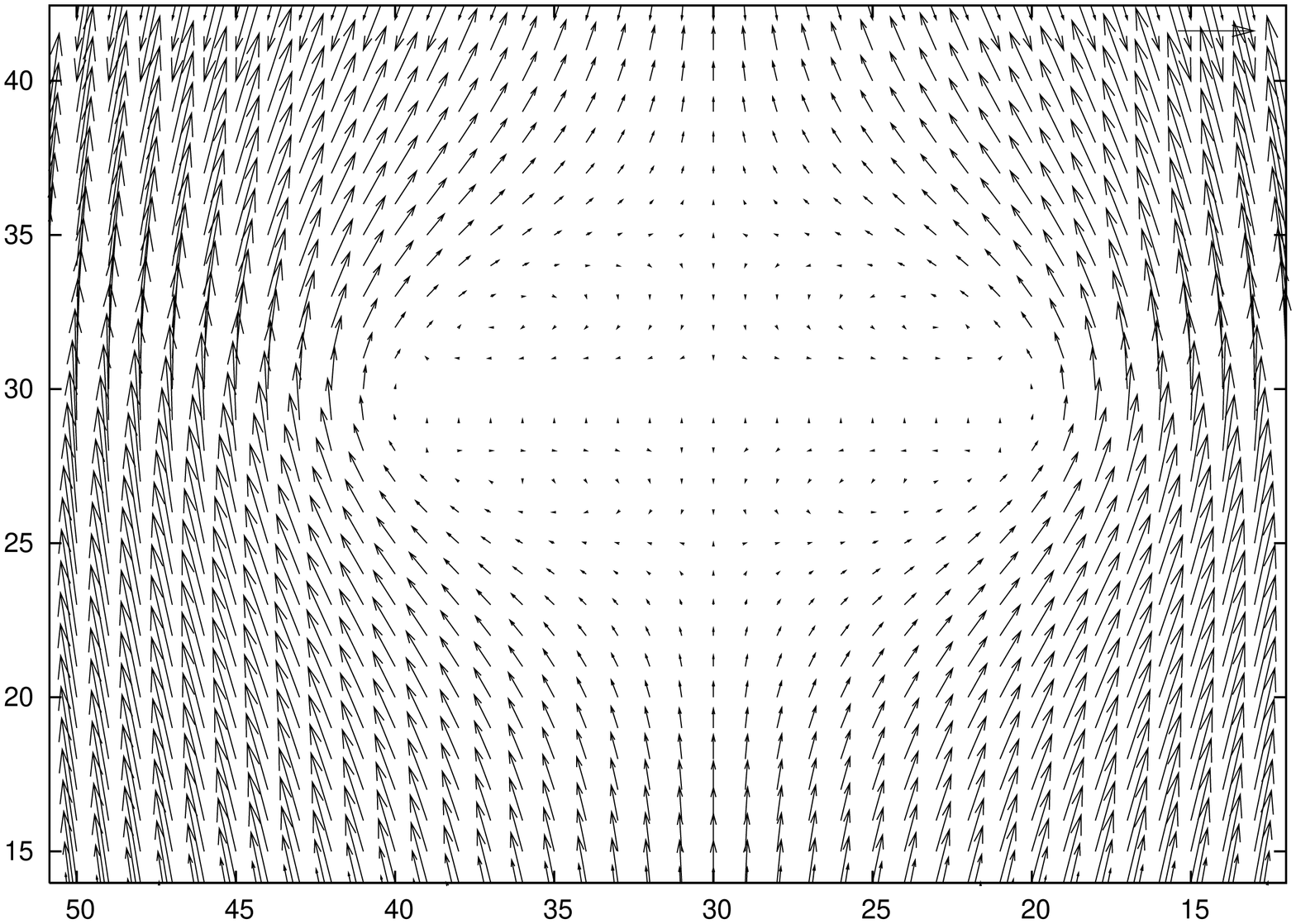}}\end{center}

\caption{Projection of the fluid velocity field near a sedimenting charged disk on
the plane parallel to the axis of revolution of the disk.
The disk sediment along its \emph{symmetry axis}.
\label{cap:Flow-broad-side}}
\end{figure}

In all figures we  observe that the reduction in sedimentation
velocity for transverse motion is larger than for longitudinal
motion. For wide double layers the differences may amount to
$20\%$. This effect can be intuitively understood in terms of the
different forces felt by the electric double layer in the two
configurations. For transverse sedimentation, most of the diffuse
layer is exposed to the flow induced by the sedimenting array of
disks. On the contrary, for longitudinal motion most of the
electric double layer is located in a region where the fluid
velocity is small and is not subject to large gradients. One would
then naively expect that this difference will be enhanced by an
increase of the surface charge. But, in fact, the relative
difference decreases with the charge of the disk. Hence, as
qualitatively illustrated in
Figs.\textbf{~}\ref{cap:Flow-edge-side} and
\ref{cap:Flow-broad-side} there are non-trivial couplings between the
electrostatic restoring force and the flow field. From the figures
it is clear that for longitudinal sedimentation most of the
diffuse layer is in a region of smoothly varying velocity, whilst
for transverse sedimentation the double-layer is exposed to large
velocity gradients.
More interestingly, by comparing the flow fields past the weakly
and the highly charged disk, we observe that the effective  hydrodynamic
shape of the particle becomes more isotropic. Moreover, close to
the surface of the particle in Fig.~\ref{cap:Flow-edge-side}b,
the direction of the flow is reversed with respect to the
direction of the bulk flow. This effect can only be caused by the
different behavior of the electrostatic and the hydrodynamic
fields at the edge of the disks, an effect we have not analyzed  in detail, because 
 much more expensive simulations are required  to gain
a more quantitative understanding of the flow patterns in
Fig.~\ref{cap:Flow-edge-side}b. Such simulations fall out of the
scope of the present paper.%

\section{Sedimentation velocity of charged disks: shape effects\label{sec: Shape}}
It is not straightforward to quantify the effect of shape
variations on the sedimentation of (charged) disks because one cannot change the shape of without modifying either the
surface charge density or the overall particle charge. Then,
because the electrostatic field next to a particle is proportional
to the surface charge $\sigma$, a change in the surface area will
change the electric field surrounding the particle, making it
impossible to isolate the effect of shape change. On the other
hand, keeping $\sigma$ constant by varying the overall particle
charge is not a solution either since the reduction in
sedimentation velocity does depend also on $Z$ {[}see
Eqn.~(\ref{eq:booth}){]}. As a result, we will have to modify both
valency and volume (to keep the surface area constant) to
disentangle charge effects from effects arising from shape
changes. However, even if we take care of this problem, we can
only compare each disk with the corresponding sphere, because the
two disks we study have different areas.

In order to focus on shape effects as much as possible, we computed
the normalized sedimentation velocity $U_{d}(Z)/U_{d}(0)$ (with $U_{d}(Z)$
the sedimentation velocity of an isolated particle with valency $Z$,
and $U_{d}(0)$ the velocity of the same object with $Z=0$) with
the corresponding normalized sedimentation velocity of a sphere with
the same valency $Z$ and surface area, $U_{s}(Z)/U_{s}(0)$.

For weakly charged particles, we can make use of Booth's
prediction to analyze the results. To this end, rather than
studying the scaled velocity directly, we have found fruitful to
consider $[1-U_{d}(Z)/U_{d}(0)]/Z^{2}$, which is the coefficient
$c_{2}$ {[}see Eqn.~(\ref{booth simple}){]} in the case of a
sphere. This is a direct measure of the electrokinetic reaction
induced by the electric double layer. Since we have argued (see
Section~\ref{sec:Charge effect}) that the charge dependence of
disks is the same as the one observed for spheres in the
Debye-H\"{u}ckel limit, the previous ratio is a quantitative way
of assessing the role of shape on the sedimentation velocity.

\begin{figure}
\begin{center}\subfigure[p=5]{\includegraphics[%
  clip,
  width=0.70\columnwidth]{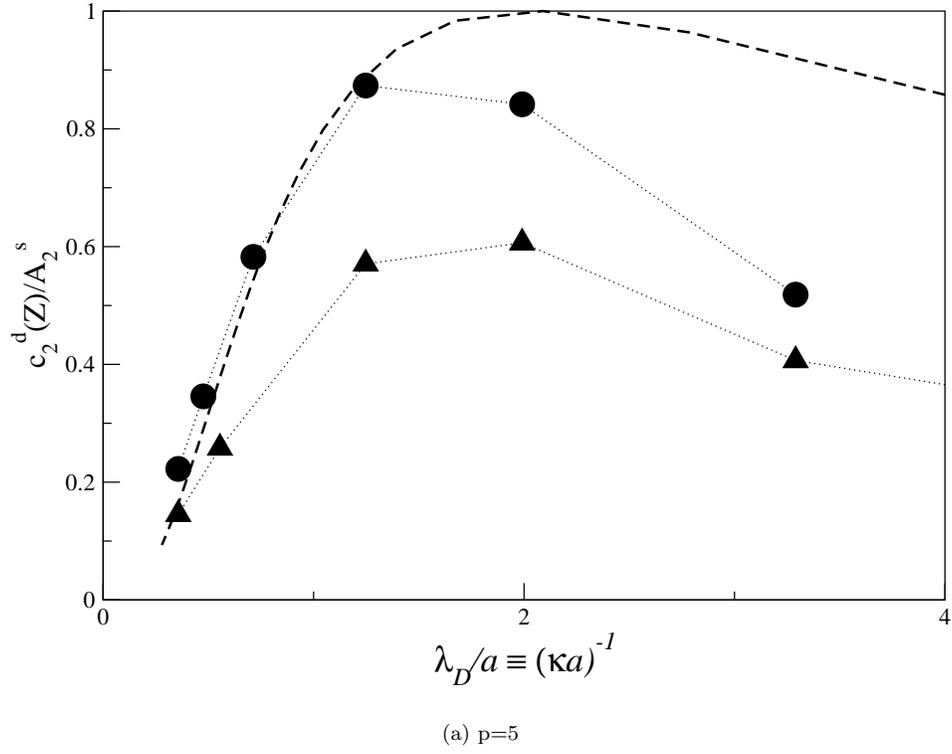}}\end{center}

\begin{center}\subfigure[p=10]{\includegraphics[%
  clip,
  width=0.70\columnwidth]{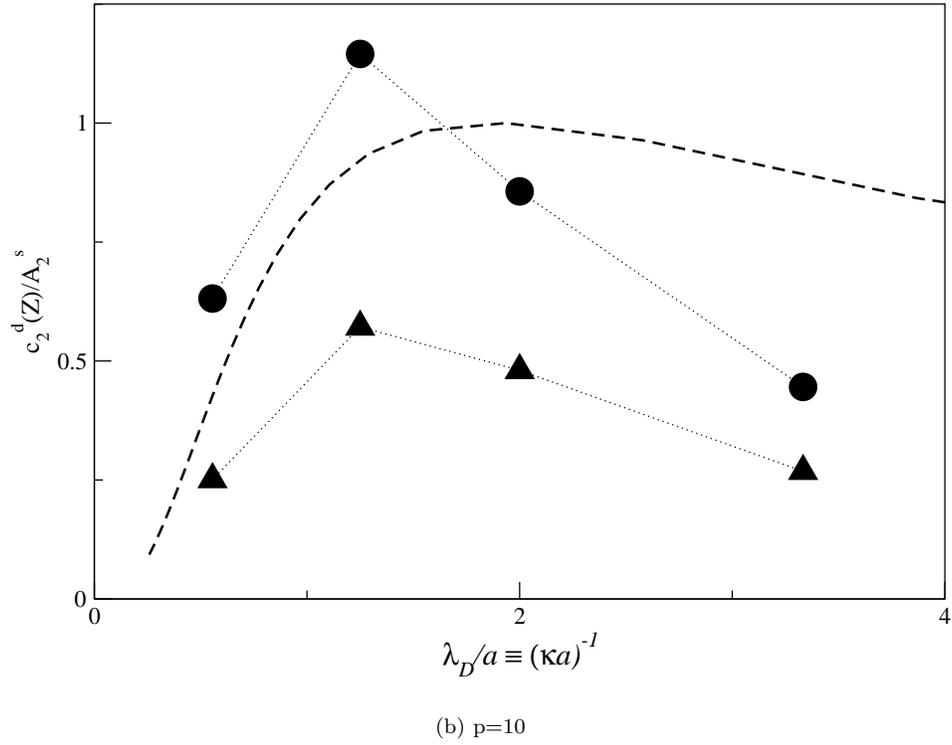}}\end{center}

\caption{Normalized reduction in the velocity of a sedimenting disk of charge $Z=10$ at
infinite dilution both for transverse (spheres) and longitudinal (squares) motion as
a function of its double layer thickness.
The dashed curve corresponds to the theoretical prediction for a sedimenting
sphere of equal charge and surface area. a) p=5, b) p=10.
Lines joining the simulation points are drawn as guide to
the eye.
\label{fig: U/U0 dilute limit.}}
\end{figure}

In Figures~\ref{fig: U/U0 dilute limit.}(a) and~\ref{fig: U/U0
dilute limit.}(b), we show $[1-U_{d}(Z)/U_{d}(0)]/Z^{2}\equiv
c_{2}^{d}$ for disks with two different aspect ratios and with a
small charge, $Z=10$, both for transverse and longitudinal
sedimentation. $c_{s}^{d}$ is expressed in units of
$A_{2}^{s}\equiv k_{B}Tl_{B}/\left(72\pi D\eta a^{2}\right)$, in
such a way that for spheres it reduces to $f(\kappa
a)$ as  predicted by Booth. For weakly charged disks and thin double
layers, the decrease in velocity does not depend strongly on shape.
This is consistent with Smoluchowski's theory for electrophoresis
\cite{vS18}, which predicts that the electrophoretic velocity of
particles with the same zeta potential (the electrostatic
potential at contact) is independent of the particle shape if
$\kappa a\rightarrow\infty$. However, the deviation from this
Smoluchowski limit appears confined at narrower double layers for
longitudinal motion; hence, shape affects significantly the
sedimentation velocity of suspended particles. Moreover, in the
case of asymmetric objects, the orientation of the particle also
affects the velocity. For both longitudinal and transverse
sedimentation, the electrokinetic coupling of a disk is always
smaller than the decrease for an equivalent sphere. One can
clearly see that the decrease in velocity for longitudinal
motion is smaller than for transverse motion. %

In the high-charge regime we use the same quantity, $c_{2}^{d}\,,$
to assess the role of shape, although we know that the Booth
theory fails in this case. In Figures~\ref{fig: U/U0 complete
p=3D5, Z=3D100}(a) and \ref{fig: U/U0 complete p=3D5, Z=3D100}(b)
we show $c_{2}^{d},$ again for two aspect ratios. In the thin
diffuse-layer limit, our data are consistent with Smoluchowski's
theory, and we observe again a departure from the results for a
sphere upon increasing the width of the electric double layer. The
maximum effect is observed for electric-double-layer widths of the
order of the largest linear dimension of the  object, and, again,
the decrease for longitudinal motion is smaller than for
transverse motion. This is consistent with the intuitive picture
that the hydrodynamic
shape of a disk becomes more isotropic upon increasing its charge. %
\begin{figure}
\begin{center}\subfigure[p=5]{\includegraphics[%
  clip,
  width=0.70\columnwidth]{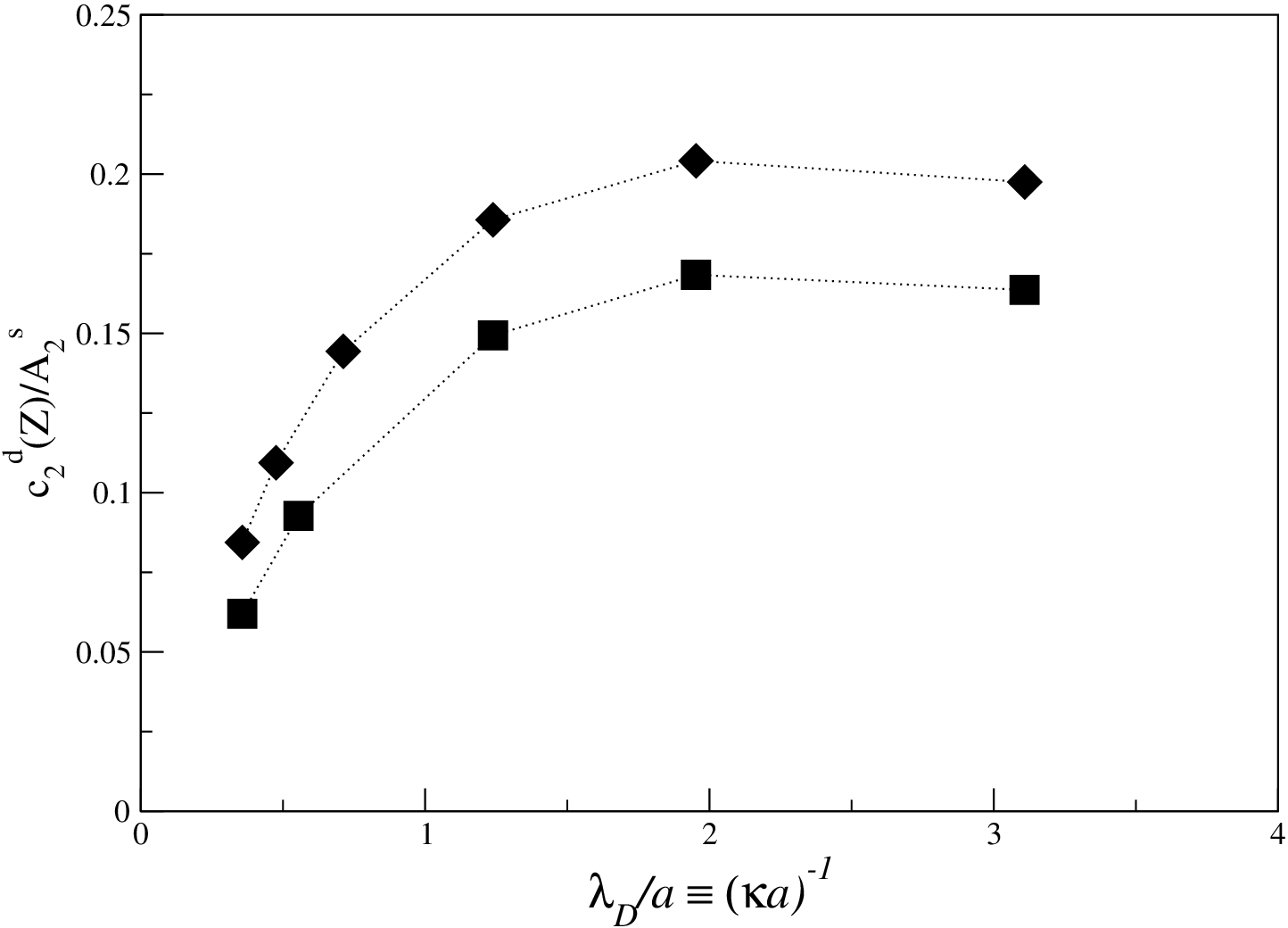}}\end{center}

\begin{center}\subfigure[p=10]{\includegraphics[%
  clip,
  width=0.70\columnwidth]{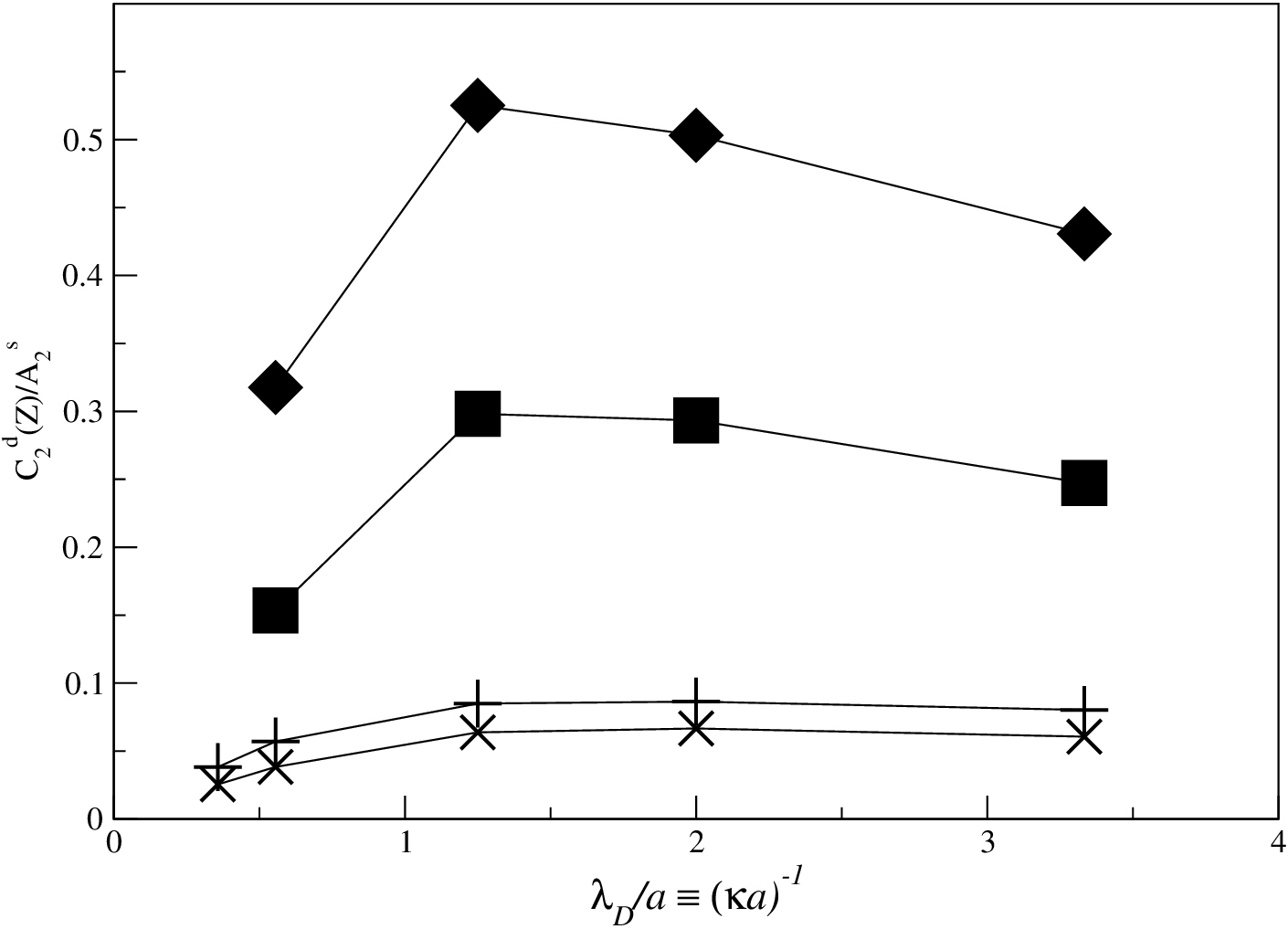}}\end{center}

\caption{
Normalized reduction in the velocity of a sedimenting \emph{highly} charged disk at
infinite dilution both for transverse (spheres) and longitudinal (squares) motion as
a function of its double layer thickness.
The dashed curve corresponds to the theoretical prediction for a sedimenting
sphere of equal charge and surface area. a) p=5, b) p=10.
Lines joining the simulation points are drawn as guide to
the eye.
Disks  and spheres have a valency $Z=100,$
which correspond to a surface charge $\sigma=0.45$ for the disk
with $p=5$ and to $\sigma=0.13$ for the disk with $p=10.$ In
Sub-figure~(b) we also show the same simulation for a disk surface
charge $\sigma=0.40$ for the disk with $p=10$ in transverse (+)
and longitudinal (x) sedimentation. Lines joining the simulation
points are drawn as a guide to the eye. \label{fig: U/U0 complete
p=3D5, Z=3D100}}
\end{figure}

By comparing Figs.~\ref{fig: U/U0 complete p=3D5, Z=3D100}(a) and
\ref{fig: U/U0 complete p=3D5, Z=3D100}(b), the reader might
conclude that the reduction in sedimentation velocity is higher
for the disk with a smaller aspect ratio. However, one should bear
in  mind that the two disks have the same valency and therefore
very different surface charges
$\sigma_{p=5}\simeq3.4\times\sigma_{p=10}.$ To show how much the
surface charge affects $c_{2}$, we show $c_{2}$ for the disk with
$p=10$ also at $Z=300.$ Even though the surface charge of this
disk is still lower than the surface charge for the other disk,
the electrokinetic effect is already more pronounced.

\begin{figure}
\begin{center}\includegraphics[%
  clip,
  width=0.80\columnwidth]{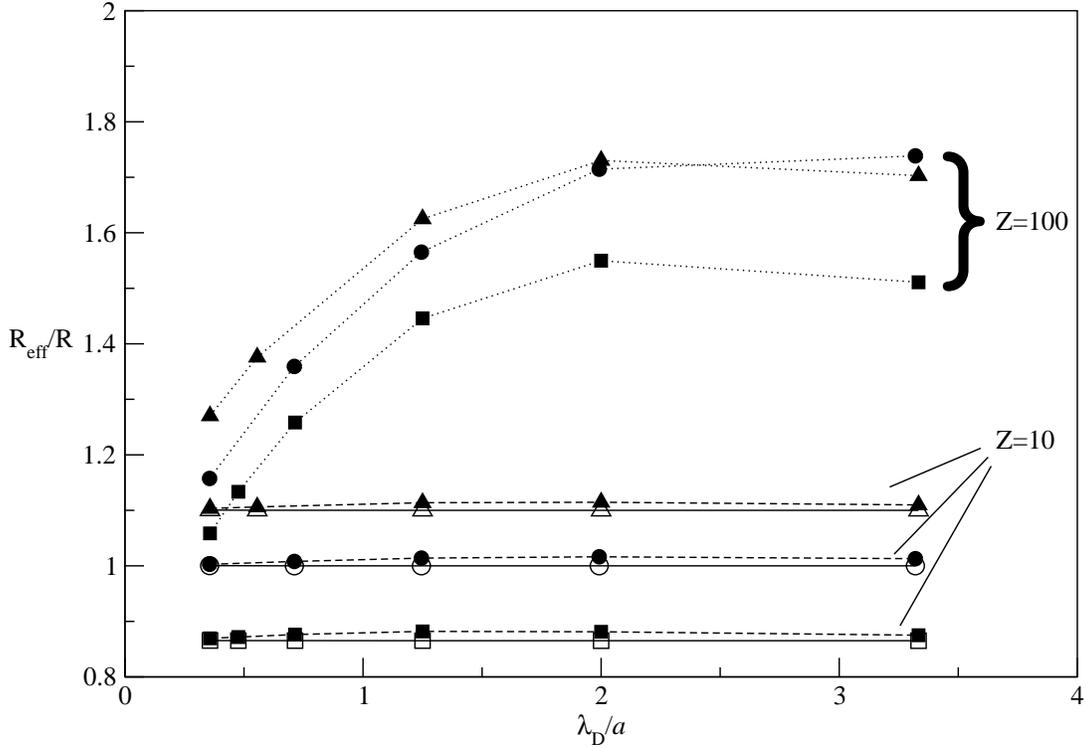}\end{center}

\caption{Effective hydrodynamic radius of a disk normalized by the
radius $R$ of a sphere with equivalent surface as a function of
the inverse Debye length in units of the disk radius. Disk with
$p=5$ in transverse (square) and longitudinal (triangle) motion;
sphere with same surface (circle). Uncharged objects (drawn line and open symbols),
$Z=10$ (dashed curve) and $Z=100$ (dotted line). The curves are
drawn as a guide to the eye. \label{fig: Effective stokes radius}}
\end{figure}

An illuminating way of displaying the relevance of shape for
sedimentation is to consider what the effective Stokes radius
of a sedimenting disk is. In Fig.~\ref{fig: Effective stokes radius}
we show the effective Stokes radius {[}$R_{{\textrm{eff}}}\equiv
F/\left(6\pi\eta U\right)${]}, where $F$ is the magnitude of the
external force acting on the disk. For small charges, the
effective radius depends weakly on the width of the double layer,
and is larger for the transverse motion, as can be expected. At
high charges the behavior is qualitatively different since
$R_{{\rm{eff}}}$ depends on the width of the double layer for
$\lambda_{D}/a<~2.$ For larger $\lambda_{D}/a$ it tends to level
off. As the diffuse layer broadens, the effective size that
characterizes the sphere and the disk in longitudinal motion tend
to converge, leading to a same effective shape for wide layers.

The physical origin of this effect is already implicit in Figs.~\ref{cap:Flow-edge-side}
and \ref{cap:Flow-broad-side}. These figures show the velocity fields
around the disk for both orientations. Different flow fields develop
around the sedimenting disk for low and high surface charge.The flow
profiles look more isotropic for high $Z,$ therefore one might expect
that for high $Z$, the friction coefficients of a disk approaches
that of a sphere with the same $Z\,.$

\section{Discussion\label{sec:Discussion}}

In this paper we have presented simulations of the sedimentation
of an array of charged disks. We have treated the electrolyte at
the Poisson--Boltzmann level, while we have incorporated the
relevant hydrodynamic couplings between the solvent and the
dissolved electrolyte. Using the lattice-Boltzmann method we have
modelled highly charged colloids and arbitrary $\kappa a$ values,
which greatly expands the parameter range that can be covered.

Since no exact analytical expressions exist for the sedimentation
velocity of isolated neutral disks and finite thickness, we have
first checked the performance of our method by validating the
sedimentation of neutral disks using approximate expressions that
become exact for infinitesimally thin disks. 
Such a computation has provided us with values for the sedimentation velocities of uncharged disks, which are needed for the subsequent analysis.

In order to clarify the role of electrohydrodynamic coupling and the
relevance of shape, we have performed a systematic study to assess
the role of shape, volume fraction, charge, and ionic strength on
the sedimentation velocity. We find that in the linearized Debye-H\"{u}ckel
regime, the sedimentation velocity has the same functional dependence
on volume fraction and surface charge as that for spheres, although with different amplitudes. 
This deviation should be accounted for when using diffusivity measurements of disks 
to infer the effective charge of colloids and this work represent, to our knowledge,
the first were the sedimentation velocity is computed systematically.
So far experimental findings could only be compared with the theory for weakly charged
spheres~\cite{TKC01}, which can lead to numerical errors in the estimates of their effective charges .  
At fixed $\kappa a,$ we have studied the surface charge dependence of the
disk sedimentation velocity, from which we have observed that in the
high-charge regime, the accumulation of charge near the disk surface
layer decreases the effect of electrokinetic coupling on the sedimentation
velocity, and also shows that such accumulation becomes more relevant
as the disk becomes more anisotropic.

We have shown that the geometrical anisotropies of neutral disks are
reduced by the presence of the electric double layer, especially for
highly charged disks. In fact, we have seen that when the double layer
is exposed to larger velocities, the reduction in sedimentation velocity
is larger. Hence, this mechanism tends to generate a more symmetric
disk response, as can be effectively characterized in terms of an
effective disk radius which becomes less sensitive to shape details
as charge increases.

\section*{Acknowledgments}
The work of the FOM Institute is part of the research program of
FOM and is made possible by financial support from the Netherlands
organization for Scientific Research (NWO).

I.P. acknowledges financial supprot from DGICYT of the Spanish Government and from
DURSI, Generalitat de catalunya (Spain), and thanks the FOM institute for its hospitality.

\bibliography{/usr/cof1/capuani/Work/Documents/fabrizio}

\end{document}